\documentstyle[12pt]{article}

\begin{document}

\title{COMETARY KNOTS AND BROAD EMISSION LINES, 
      GAMMA RAYS AND NEUTRINOS FROM AGN} 
\author{Arnon Dar\\
Department of Physics and Space Research Institute\\
Technion-Israel Institute of Technology \\
Haifa 32000, Israel\\}

\maketitle 

\centerline{Abstract}

Recent observations with the Hubble Space Telescope have discovered in the
nearest planetary nebula, the Helix Nebula, thousands of gigantic
comet-like objects with planet like masses in a ring around the central
brilliant star at a distance comparable to our own Oort cloud of comets.
We propose that such circumstellar rings of planets exist around most
stars and that the gas clouds which emit the broad optical lines from
quasars are radiation ablated planets which have been stripped off by
gravitational collisions from stars that orbit near the central massive
black hole. We propose that collisions of jet accelerated particles with
these targets crossing the line of sight produce TeV $\gamma$-ray
flares (GRFs) from blazars like Markarian 421 and Markarian 501. Hadronic
production of TeV GRFs from blazars implies that they are accompanied by a
simultaneous emission of high energy neutrinos, and of electrons and
positrons with similar intensities, light curves and energy spectra.
Cooling of these electrons and positrons by emission of synchrotron
radiation and inverse Compton scattering produces $\gamma$-ray, X-ray,
optical and radio afterglows. 

\section{Introduction} 

In a recent letter we have suggested (Dar and Laor 1997) that collisions
of jet accelerated particles with gas clouds in the broad line emission
region of Active Galactic Nuclei (AGN) close to the line of sight produce
TeV $\gamma$-ray flares (GRFs) from blazars like Markarian 421 (Mrk 421)
and Markarian 501 (Mrk 501). In this letter we suggest that these gas
clouds are ablated planets, like the $\sim 3500$ gigantic ``Cometary
Knots'' which were discovered with the Hubble Space Telescope (O'Dell and
Handron 1996) in a ring around the central star in the Helix nebula.  We
suggest that distant circumstellar rings which contain thousands of
planets may exist around most stars. Such losly bound planets are stripped
off (``planetization'') from stars which orbit near the central massive
black holes in AGN by gravitational collisions. They are ablated by the
strong radiation field and by gravitational collisions and form the
ionization bounded ``gas clouds'' which emit the broad optical lines from
AGN. We also suggest that relativistic jets in AGN act as Fermi magnetic
mirrors which accelerate the ionized interstellar atoms in front of the
jets to cosmic ray energies..  In blazars, which are AGN with relativistic
jets that happen to lie near the line of sight, strong GRFs are produced
when Cometary Knots with high column density cross the line of sight near
the central black hole. TeV $\gamma$-rays are then produced through the
reaction $pp\rightarrow \pi^0 X$ which is followed by prompt
$\pi^0\rightarrow 2\gamma$ decay. The production of TeV $\gamma$-rays is
also accompanied by production of TeV neutrinos, electrons and positrons,
mainly via $pp\rightarrow \pi^{\pm}X$; $\pi^{\pm}\rightarrow
\mu^{\pm}\nu_\mu$;  $\mu^{\pm}\rightarrow e^{\pm}\nu_e\nu_\mu$.  The
subsequent cooling of these electrons and positrons by synchrotron
radiation, inverse Compton scattering, bremmstrahlung and annihilation
produces $\gamma$-ray, X-ray, optical, and radio afterglows. We show that
this model reproduces well the main observed properties of GRFs from Mrk
429 and Mrk 501 (Punch et al. 1992; Lin et al. 1992; Macomb et al. 1995; 
Schubnell et al. 1996; Gaidos et al. 1996; Quinn et al. 1996; Catanese et
al. 1997;  Bradbury et al. 1997; Schubnell 1997).

\section{Thousands of Planets In Circumstellar Rings?}

Recent observations with the Hubble Space Telescope of the Helix Nebula,
the nearest planetary nebula at an estimated distance of 150 pc, have
discovered that the central star lights a circumstellar ring of about 3500
gigantic comet-like objects (``Cometary Knots'') with a typical mass of
$M_c \sim 1.5\times 10^{-5}M_\odot$ (O'Dell and Handron 1996) comparable
to our solar system planets ($M_{Earth}=3\times 10^{-6}M_\odot$,
$M_{Jup}=9.6\times 10^{-4}M_\odot$) and a total mass of $\sim
10^{-2}M_\odot$. The gaseous heads of these Cometary Knots have a typical
size of about $r_c\sim 10^{15}cm$ and an average column density of
$M_c/\pi m_p r_c^2\sim 10^{22} cm^{-2} $.  They look like ionization
bounded neutral clouds, but it is not clear whether they contain a solid
body or uncollapsed gas.  They are observed at distances comparable to our
own Oort cloud of comets, but they seem to be distributed in a ring rather
than in a spherical cloud like the Oort cloud:  The Helix Nebula has a
structure of a ring lying close to the plane of the sky. The Cometary
Knots are not seen close to the central star, but become very numerous as
one approaches the inner ring and then become fewer out in the main body
of the ring (perhaps, because it becomes more difficult to detect them
there against the bright background).  This suggests a ring rather than a
spherical distribution, since no foreshortened objects close to the
direction of the central star are seen. It is possible that the Cometary
Knots have been formed together with the central star (O'Dell and Handron
1996) since star formation commonly involves formation of a thin planar
disk of material possessing too high an angular momentum to be drawn into
the nascent star and a much thicker outer ring of material extending out
to several hundreds AU (Beckwith 1995). Evidence for this material has
been provided by infrared photometry of young stars and also by direct
imaging of this material (O'Dell and Wen 1994). The observation of comets
of very long period and unconfined to the ecliptic plane gave rise to the
idea of the Oort cloud of billions of comets in our Solar System at
distances out to many tens of thousands of AU (Oort 1950). It is argued
that small gravitational perturbations have circularized their parking
orbits where occasionally another perturbation puts them into an orbit
which brings them into the inner planetary system where they are finally
viewable as comets. The mass distribution of the observed comets has been
determined over only a small range since there are observational selection
effects acting against finding the small ones and there are very few large
ones.  It is possible that the Cometary Knots are the high mass end of the
vastly more numerous low mass comets (O'Dell and Handron 1996) and are
confined to the ecliptic plane because of their relatively larger masses. 
In this paper we propose that most stars are surrounded with circumstellar
rings which contain thousands of planets.  These loosely bound planets are
stripped off from their mother stars by gravitational collisions in dense
stellar regions. They are ablated by collisions and the strong radiation
field near the massive central black hole in AGN and form their Broad
Emission-Line Region (BLR).

\section{Broad Emission-Lines Clouds In Quasars} 

Detailed studies of broad optical and ultraviolet emission-lines, whose
atomic physics is well understood, have been used to obtain detailed
information on the BLR of AGN. From their line-shapes, relative strengths
and their time-lag response to the variations with time of the central
continuum source, it was concluded that the BLR consists of a very large
number $>10^6$ of ionization bounded neutral clouds that move with very
large random velocities and produce the broad emission-lines. The column
density and mean density of the broad emission-line clouds and
consequently their mean size, were estimated from the ionizing flux of the
central source and the relative line strengths from the partially ionized
clouds. Quite high densities and column densities were inferred. Typical
values are, $N_p\sim 10^{22-25}~cm^{-2}$ and $n_p\sim 10^{10-12}~cm^{-3},$
respectively.  For spherical clouds of uniform density, $\bar N_p=(4/3)n_p
r_c$. Consequently, the radii of the clouds are typically , $r_c\sim
10^{13\pm1}cm$. The velocity distribution of the clouds has been estimated
from the profiles of the broad emission lines. Their full widths at half
maximum indicate typical velocities of a few $10^3~ km~s^{-1}$ extending
beyond $10^4~km~s^{-1}$ at the base of the lines. Reverberation mapping
has clearly established that the velocities are not a radial flow (Maoz
1997).  The size of the BLR has been estimated from reverberation mapping
of both Seyfert 1 galaxies (e.g., Peterson 1993) and quasars (e.g. Maoz,
1997), with typical lag times between 10 days for Seyfert 1 galaxies and
100 days for quasars, respectively. Typically, $R_{BLR}\approx 3\times
10^{16}L_{44}^{1/2}~ cm,$ where $L=L_{44}\times 10^{44}~erg~s^{-1}$ is the
luminosity of the AGN in ionizing radiation. The velocities of the clouds
seem to be consistent with those expected for clouds orbiting massive
black holes, $v_c\approx \sqrt {GM/R}\approx 1.15\times 10^9
\sqrt{M_8/R_{16}}~cm~s^{-1}~,$ where $M=M_8\times 10^8M_\odot$ is the mass
of the black hole and $R=R_{16}\times 10^{16}~cm$ is the distance from the
black hole.  The total number of clouds in the BLR was estimated from the
sizes of the BLR and clouds and the covering factor $C$, i.e., the
fraction of the AGN sky covered by clouds. The latter was estimated from
the ratio of Ly$\alpha$ photons emitted by the clouds to the H ionizing
photons produced by the central continuum. Typically, $C\sim 0.1~.$
Assuming $C\ll 1$, one finds $N_c=(4/3) C R^2_{BLR}/r^2_c~$. i.e.,
typically, $N_c\sim 10^6-10^9$. This number is consistent with the number
of Cometary Knots expected in a galactic core like that of our Milky Way
galaxy with typically $\sim 10^6$ stars within its inner 1 pc and $\sim
10^3$ per star.

\section{Relativistic Jets As Cosmic Accelerators} 

GeV $\gamma$-rays have been detected from more than 50 active galactic
nuclei (AGN) by the Energetic Gamma Ray Telescope Experiment (EGRET) on
the Compton Gamma Ray Observatory (e.g., von Montigny et al. 1995;
Thompson et al. 1995), all belonging to the blazar type of AGN. TeV gamma
ray emissions have been observed only from the two nearest BL Lac objects,
Mrk 421 at redshift z=0.031 and Mrk 501 at redshift z=0.033 (Punch et al.
1992; Lin et al. 1992; Macomb et al. 1995;  Gaidos et al 1996, Quinn et al
1996; Schubnell et al. 1996; 1997).  But, it has been suggested that
perhaps all $\gamma$-ray blazars emit TeV $\gamma$-rays and the opacity of
the intergalactic space to TeV photons due to $e^+e^-$ pair production on
infrared background photons prevents us from seeing them in TeV photons
(e.g., Stecker et al. 1993). 

The observed emission of GeV and TeV $\gamma$-rays requires a power-law
spectrum of accelerated particles which extends to very high Lorentz
factors, much higher than the typical bulk Lorentz factors, $\Gamma_b\leq
10$, of AGN jets (see, e.g., Guerra and Daly 1997). The observed GeV and
TeV $\gamma$-rays from blazars must be emitted from the relativistic jets
far out from the central object. This is because near the central object
acceleration to very high Lorentz factors is prevented by fast inverse
Compton cooling of electrons which are coupled by Coulomb interactions to
protons. It is also required in order to avoid self absorption by $e^+e^-$
pair production. It is still not known how such relativistic jets are
formed in AGN, what is their particle composition, and how they accelerate
particles to very high energies with a power law spectrum. Here we outline
a simple mechanism by which relativistic jets with bulk Lorentz factors of
only $\leq 10$ may accelerate particles to very high energies with a
power-law spectrum $dn/dE\sim E^{-\alpha}$ where $\alpha\sim 2.5$. 
 
The high collimation of AGN jets over huge distances (up to hundreds of
kpc), the confinement of their highly relativistic particles, their
emitted radiations and observed polarizations, all indicate that AGN jets
are highly magnetized, probably with a strong helical magnetic field along
their axis. The UV light emitted from the AGN and the jet ionizes the
interstellar medium (ISM) in front of the jet.  The jet magnetic field
then acts as a magnetic mirror and accelerates the ionized interstellar
particles to highly relativistic energies through the Fermi mechanism
(1949): 

In the rest frame of a jet, which moves with a bulk Lorentz factor 
$\gamma_b$, the charged ISM particles are moving towards
the jet with an energy $\gamma_b mc^2$ and are reflected back with the
same energy by the transverse magnetic field in the jet. In the observer
frame their energy is boosted to $E=\gamma_b^2 mc^2$. Each time such a
charged particle is reflected by an external magnetic field (of the ISM or
a star) back into the jet its energy is boosted further by a factor
$\gamma_b^2$. This efficient acceleration continues until the jet becomes
non relativistic. Because of the small cross sections for binary
collisions of relativistic particles and the low Lorentz factor of the jet
bulk motion, the jet looses most its energy by acceleration of the ISM and
not by direct radiation or by binary collisions with the ISM particles. 

High resolution observations in radio (VLB and VLBI) and in optical (HST)
wavelengths indicate that jet ejection from AGN and microquasars is not
continuous but occurs in ejection episodes. Let us denote by M the total
ejected mass in an ejection episode, by $\Gamma_B$ its bulk Lorentz factor
and by $M_{ISM}$ the total mass of ionized ISM which is accelerated by M.
For simplicity let us assume a pure hydrogenic composition (the
generalization to an arbitrary composition is straight forward).
Conservation of energy and momentum then reads approximately
\begin{equation}
                   d(Mc^2\gamma_b)\approx -d(M_{ISM}) c^2\gamma_b^2.   
\end{equation} 
Consequently, for an ISM with a uniform composition 
\begin{equation}
    {dn_p\over dE}\approx {M\over 2m_p} \left[{E\over m_p}\right ]^{-3/2}; 
                           ~~~E<\Gamma_B^2mc^2. 
\end{equation} 
Note that the $\sim E^{-3/2}$ power-law spectrum is independent of whether
the jet has a conical geometry (transverse expansion) or a cylindrical
geometry (transverse confinement) and it is the same for ions and
electrons. The composition of the accelerated particles reflects the
composition of the ionized matter in the ISM. If one assumes that the
probability of an accelerated ion to be reflected back (by ISM or stellar
magnetic fields) into the jet 
decreases like $R_L\sim E^{-1}$ where $R_L$ is the Larmour radius, and
that it is independent of ion type (same charge to mass ratio), then repeated
acceleration will produce a power-law spectrum,
\begin{equation}
                   {dn_p\over dE}\approx{\Gamma_BM\over 2m_p}\left [{ 
E\over m_p}\right ]^{-5/2}. 
\end{equation} 
This power-law spectrum of the accelerated particles cuts off when the
Larmour radius of the accelerated particles ceases to be small compared
with the size of the ionized region and the range of the relativistic jet. 
One can check that the above picture yields an estimated range $L$ of AGN 
jets that is consistent with observations:  Let us assume that a fraction
$\eta$ of the accretion power $\dot{M}c^2$ is continuously converted into a
relativistic kinetic energy of a jet which is injected into a solid angle
$d\Omega = \pi\theta^2\ll 4\pi$. Let $c_s\approx \sqrt {5kT/3m_p}$ be the
sound speed in the ISM, i.e., the speed of ISM material which replaces the
accelerated material. The jet propagates to a distance where it becomes
non relativistic and the time it takes it to reach there becomes longer
than the time it takes the ISM to replace the accelerated mass by new
mass, $\eta \dot{M} L\theta/c_s \approx n_p m_p \pi\theta^2L^3/3$.
Consequently,
\begin{equation}   
         L=\left({3\eta\dot{M}\over c_s n_p m_p \pi\theta}\right)^{1/2}
\end{equation}   
For typical parameters, $\eta \dot{M}=M_\odot~y^{-1}$, $T=10^5K$,
$\theta\approx 3^0$, one obtains $L\approx 10~ kpc$ for
$n_p=1~cm^{-3}$ and $L\approx 200 kpc$ for $n_p=10^{-3}~cm^{-3}$,
respectively. 
 
\section{The Hadronic Collider Model For Blazars} 
Following Dar and Laor
(1997) we propose that TeV $\gamma$-rays which are beamed towards the
observer are produced through $pp\rightarrow \pi^0X$; $\pi^0\rightarrow
2\gamma$ when ``Cometary Knots'' from the BLR cross the jet near the line
of sight. The quiescent emission is due to synchrotron emission and
inverse Compton scattering of external photons or photons emitted from the
jet. The quasi-quiescent emission is due to jet interactions with many,
relatively distant, ``Cometary Knots'' in the BLR. Strong GRFs are
produced when Cometary Knots with large area of high column density cross
the line of sight at relatively small distance from the central engine. 
The hadronic collider model for blazars predicts TeV $\gamma$-ray emission
which strongly fluctuates with time and shows spectral evolution, even if
jet ejection does not vary with time on short time scales.  The exact
properties of individual flares depend on many unknown parameters of both
the Cometary Knots (their geometry, density distribution, speed and
trajectory relative to the jet and line of sight) and the jet (geometry,
exact orientation relative to the observer, particle composition and
differential energy spectrum of its high energy particles as function of
distance from the jet axis and along the jet). However, the general
properties of the quasi-quiescent emission and the flares can be estimated
using some simplifying assumptions. Essentially, the results of the
hadronic collider model of blazars (Dar and Laor 1997) are applicable here
with the only substitution ``clouds = Cometary Knots''.  We summarize them
briefly below: 

Because of the exponential dependence of the production cross section for
$pp\rightarrow\gamma X$ on the transverse energy of the produced
$\gamma$-rays (Neuhoffer et al. 1971; 
Boggild and Ferbel 1974; Ferbel and Molzon 1984), most of
the $\gamma$-rays, which are seen by the observer when a gas cloud crosses
the line of sight at a distance R from the central black hole,
must arrive from impact parameters smaller than the critical impact
parameter $b_c \approx RE_0/E_\gamma<R\theta_{jet}$, where $E_0$ is the
average transverse energy of the produced $\gamma$-rays ($E_0\sim
0.16~GeV$ is independent of incident energy).  The number of clouds with
$b<b_c$ in the BLR is $N_cE_0^2/4E_\gamma^2.$ A quiescent background is
formed by jet-cloud interactions only if this number is large, i.e., if
$E_\gamma\ll E_{crit}\approx\sqrt{N_c}E_0/2\approx \sqrt{C_{0.1}L_{44}}
/r_{12}~TeV,$ where $r_c=r_{12}\times 10^{12}cm$. In that case the
jet produces a quiescent $\gamma$-ray flux of
\begin{equation} 
{dI_\gamma\over dE_\gamma}
      \approx C \bar{N}_p\sigma_{in}IAE_\gamma^{-\alpha}.
\end{equation} 
and the BLR acts as a target with an effective column density
of~$C\bar{N}_p$, as long as $E_\gamma>E_0/\theta_{jet}$ (below this energy
the produced $\gamma$-rays are not beamed effectively towards the
observer). For $E_\gamma > E_{crit}$ the BLR emission is expected to
fluctuate considerably. A strong flare 
relative to the quasi-quiescent background is formed when a
cloud with a high column density crosses the line of sight at relatively a
small $R$.  If the radius $r_{eff}$ of the area with a high column density
is larger than the critical impact parameter, i.e.,
$r_{eff}>RE_0/E_\gamma$, then when the cloud blocks the line of sight, the
$\gamma$-ray flux at photon energies $E_\gamma>(R/r_{eff})E_0\sim
1.6R_{16}/r_{12}~ TeV$ flares up with a maximum intensity,
\begin{equation} 
{dI_\gamma\over dE_\gamma}
      \approx  N_{eff}\sigma_{in}IAE_\gamma^{-\alpha}, 
\end{equation} 
where 
\begin{equation}
N_{eff}\approx \int_0^{b_c} d^2b[1-e^{-\sigma_{in}N_p}]/\pi b_c^2\sigma_{in}
\end{equation}
is the effective column density.    
Thus, the maximal intensity contrast  of TeV  GRFs  compared with   
the quiescent emission is $N_{eff}/C\bar{N}_p\approx 10-100 $.  
The total duration of strong TeV emission in such flares is of the order of 
the time it takes the core of the cloud to cross  the line of sight, i.e.,
\begin{equation}
T\sim r_{eff}/v_c \sim  10^3 r_{12}R_{16}^{1/2}M_8^{-1/2}~ s.
\end{equation}  
The mean time between such strong flares is 
\begin{equation}
\Delta t\approx 
(R_{BLR}E_0/Cr_{eff}E_\gamma )T(E_\gamma)\approx
50 L_{44}^{1/2}/C_{0.1}E_{TeV}r_{12}]T  
\end{equation}
where $C=0.1C_{0.1}$. 
For $E_\gamma< 1.6 R_{16}/r_{12}~TeV$ the maximal GRF intensity is reduced 
by $(r_{eff}E_\gamma/RE_0)^2$ and the duration of the GRF is approximately
the time it takes the cloud to cross the beaming cone:
\begin{equation}
T \sim RE_0/v_cE_\gamma\sim 1.4\times 
10^{6}(R_{16}^{3/2}/E_{GeV}^{-1}M_{8}^{-1/2})~ s.  
\end{equation} 
Hence the GRF has the following general behavior when a cloud crosses the
line of sight at a distance $R$:  At energies well below $E_\gamma\sim
1.6 R_{16}/r_{12}~TeV$, the intensity contrast increases with
increasing energy while the duration becomes shorter. Above this energy
both the intensity contrast and the duration become independent of energy.
This behavior results in a spectrum which becomes harder when the
intensity increases and softens when the intensity decreases. The averaged
quasi-quiescent emission spectrum therefore is softer than the spectrum of
strong flares at peak intensity. 

The above predicted properties of the quiescent emission and the flaring
of blazars in TeV $\gamma$-rays seem to be supported by the
observed properties of TeV flares and quasi-quiescent emission from
Mrk 421 and Mrk 501 (Punch et al. 1992;  Lin et al. 1992; Kerrick et al.
1995; Macomb et al. 1995; Quinn et al. 1996; Gaidos et al 1996;
Schubnell et al. 1996; Catanese et al. 1997). 

\subsection{Production Of Neutrinos} 

Hadronic production of photons in diffuse targets is also accompanied by
neutrino emission through $pp\rightarrow\pi^{\pm}X$; $\pi^{\pm}~\rightarrow
\mu^{\pm}\nu_\mu$; $\mu^{\pm} \rightarrow e^{\pm}\nu_\mu\nu_e $.  If the
incident protons have a power-law energy spectrum, $dF_p/dE=
AE^{-\alpha}$, and if the cloud is transparent both to $\gamma$-rays and
neutrinos, then because of Feynman scaling, the produced high energy
$\gamma$ rays and neutrinos have the same power law spectrum and satisfy
(e.g., Dar and Shaviv 1996):
\begin{equation} 
dI_\nu/dE\approx 0.7 dI_\gamma/dE\sim E^{-\alpha}.
\end{equation}
Consequently, we predict that $\gamma$-ray emission from blazars is
accompanied by emission of high energy neutrinos with similar fluxes,
light curves and energy spectra. The number of $\nu_\mu$ events from a GRF
in an underwater/ice high-energy $\nu_\mu$ telescope is $SN_AT_{GRF}\int
R_\mu(d\sigma_{\nu\mu}/dE_\mu)(dI_\nu/ dE)dE_\mu dE$, where $S$ is the
surface area of the telescope, $N_A$ is Avogadro's number,
$\sigma_{\nu\mu}$ is the inclusive cross section for $\nu_\mu p
\rightarrow \mu X$, and $R_\mu$ is the range (in $ gm~cm^{-2}$) of muons
with energy $E_\mu$ in water/ice. For a GRF with $I_\gamma \sim
10^{-9}~cm^{-2}s^{-1}$ above $E_\gamma=1~TeV$ and a power index $\alpha=2$
that lasts 1 day, we predict 3 neutrino events in a $1~km^2$ telescope.
Since the universe is transparent to neutrinos, they can be used to detect
TeV GRFs from distant $\gamma$-ray blazars. If the reported GeV GRF from
the brightest EGRET $\gamma$-ray blazar PKS 1622-297, which had a maximal
flux of $I_\gamma\sim 1.7\times 10^{-5} ~cm^{-2}s^{-1}$ photons above 100 MeV
(Mattox et al 1997), was accompanied by a TeV GRF it could have produced
$\sim 30~\nu_\mu$ events within a day in a $1~km^2$ neutrino telescope.. 

\section{X-Ray, MeV and GeV GRFs} 
Hadronic production of TeV $\gamma$-rays is also
accompanied by production of TeV electrons and
positrons mainly via $pp\rightarrow \pi^{\pm}X$; $\pi^{\pm}\rightarrow
\mu^{\pm}\nu_\mu$;  $\mu^{\pm}\rightarrow e^{\pm}\nu_e\nu_\mu$. 
Their production suddenly enriches the jet with high energy electrons. Due to
Feynman scaling, their differential spectrum is proportional to the
$\gamma$-ray spectrum
\begin{equation} 
dI_e/dE\approx 0.5 dI_\gamma/dE
\end{equation}
and they have the same power-index $\alpha$ as that of the 
incident protons and the produced high energy photons and neutrinos.
Their cooling via synchrotron
emission and inverse Compton scattering from the internal (jet)
and external (cloud) magnetic and radiation fields, respectively, produce 
delayed emission of $\gamma$-rays, X-rays, optical photons and radio waves
with a differential power-law spectrum (assuming no absorption in the 
cloud)
\begin{equation}
dI_\gamma/dE\sim E^{-(\alpha+1)/2},  
\end{equation}  
where $(\alpha+1)/2\approx 1.75\pm 0.25~.$ Hence,  emission of TeV
$\gamma$-rays is accompanied by delayed emission (afterglows) in the
$\gamma$-ray, X-ray, optical and radio bands. 
 
The peak emission of synchrotron radiation by electrons with a
Lorentz factor $\Gamma_e$ traversing a perpendicular magnetic field
$B_\perp(Gauss) $ in the jet rest frame which moves  with a 
bulk Doppler 
factor $\delta=(1-\beta
cos\theta)/\Gamma$ occurs at photon energy (Rybicki and 
Lightman 1979) $E_\gamma \sim 5\times
10^{-12} B_\perp\Gamma_e^2\delta ~keV $. The electrons lose $\sim
50\%$ of their initial energy by synchrotron radiation in
\begin{equation}
\tau_c\approx 5\times 10^8 \Gamma_e^{-1} B_\perp^{-2}\delta~s\approx 
1.2\times 10^3B_\perp^{-3/2}E_\gamma^{-1/2}\delta^{-1/2}~s. 
\end{equation} 
Consequently, the time-lag of synchrotron emission is inversely
proportional to the square root of their energy. It is small for
$\gamma$-rays and X-rays but considerable ($\sim hours$) for optical
photons and ($\sim$ days) for radio waves.  The time variability of the
intensity in different energy bands decreases with frequency. The
integrated afterglow energy over the radio-optical-X-ray and $\gamma$-ray
bands is limited by the total electron energy to less than $\sim 50\%$ of
the total energy in the TeV GRF. The spectral evolution of the afterglow
is a convolution of the spectral evolution of the production of high
energy electrons and their cooling time. It is hardest around maximum
intensity and softens towards both the beginning and the end of the flare.
Because of electron cooling the spectrum should be harder during rise time
than during decline of the flare. Such feature features seem to have been
observed by ASCA (Takahashi et al. 1996) in the X-ray flare (XRF) that
followed the TeV GRF from Mrk 421 on May 15, 1995. 

\section{Discussion and Conclusions}

The observed GeV and TeV $\gamma$-ray emissions from blazars are usually
interpreted as produced by inverse Compton scattering of highly
relativistic electrons in the jet, on soft photons, internal or external
to the jet, (e.g., Maraschi et al. 1992, Bloom and Marscher 1993, Dermer
and Schlickeiser 1993; 1994, Coppi et al. 1993: Sikora et al. 1994;
Blandford and Levinson 1994; Inoue and Takahara 1996).  Although quiescent
radio, X-ray and $\gamma$-ray emissions are naturally explained by
synchrotron radiation and inverse Compton scattering of high energy
electrons in the jet, there are inherent difficulties in explaining TeV
$\gamma$-ray emission as inverse Compton scattering of soft photons by
highly relativistic electrons or positrons in pure leptonic jets. The main
difficulty is the fast cooling of electrons and positrons by inverse
Compton scattering in the very dense photon field near the AGN (Levinson
1997), when they are accelerated to the very high energies required for
the production of multi TeV $\gamma$ rays. Moreover, such a model does not
provide a natural explanation for the very short time-scale variability of
the emitted high energy radiations (because of limited statistics the
exact short time variability is probably not known yet). In this letter we
have proposed an alternative model for TeV emission from blazars based on
the assumption that AGN jets accelerate normal hadronic matter (e.g.,
Mannheim and Bierman 1992).  TeV $\gamma$-rays are produced efficiently by
the interaction of the high energy protons accelerated by the jet with gas
targets of sufficiently large column density that cross the jet. Such gas
targets could have been stripped off from the circumstellar rings around
stars that orbit near the central black hole. Thousands of such gigantic
cometary like objects have been discovered recently with HST in a ring
around the central star in the Helix nebula. Their properties and number
can account for the broad emission line clouds in AGN.  The simple
properties of hadronic production of high energy $\gamma$-rays, which are
well known from lab experiments, together with the properties of the
gigantic cometary knots can explain both the observed quasi-quiescent
emission and the TeV $\gamma$-ray flares from blazars. We also predict
prompt emissions of TeV neutrinos with comparable fluxes and delayed
emission (afterglows) from GRFs in the $\gamma$-ray, X-ray, optical and
radio bands with comparable integrated energies. Detailed predictions
depend on many unknown parameters, but many general predictions do not
depend on their choice and they seem to agree with the observations of
high energy $\gamma$-ray emissions from Mrk 421 and Mrk 501. They seems to
support an hadronic origin of TeV $\gamma$-rays emission from blazars.
Although further observations of TeV $\gamma$-ray emission and other
emissions from blazars may provide more supporting evidence for the
hadronic nature of AGN jets, a decisive evidence will probably require the
detection of TeV neutrino fluxes from Blazars, perhaps by the proposed 1
km$^3$ neutrino telescopes. 

\noindent 
{\bf Acknowledgment}: The authors would like to thank A. Laor for
useful discussions.

\centerline{{\bf References}}   

\noindent 
Beckwith, S.V.W., 1995, in {\it Science with the VLT}, eds.
J. R. Walsh and T. J. Danziger (springer, Heidelberg), p. 53 

\noindent
Blandford, R. D. \& Levinson, A. 1995, ApJ, {\bf 441}, 79 \hfill

\noindent
Bloom, S. D. \& Marsher, A.P. 1993, AIP {\bf 280}, 578 \hfill

\noindent
Boggild, H. \& Ferbel, T. 1974, Ann. Rev. Nucl. Sci. {\bf 24}, 451 \hfill

\noindent
Bradbury, S. M. et al. 1997, A\&A, {\bf 320}, L5  

\noindent
Catanese, M. et al. 1997, preprint, astro-ph 9707179

\noindent
Coppi, P. S., et al., 1993, AIP {\bf 280}, 559 \hfill

\noindent
Dar, A. \& and Laor, A. 1997, in preparation   

\noindent
Dar, A. \& Shaviv, N. 1996, Ap. Phys. {\bf 4}, 343 

\noindent
Dermer, C.D. \& Sclickeiser, R. 1993, ApJ, {\bf 415}, 418

\noindent
Dermer, C.D. \& Sclickeiser, R. 1994, ApJS, {\bf 90}, 945

\noindent
Ferbel, T. \& Molzon, W.R. 1984, Rev. Mod. Phys., {\bf 56}, 181

\noindent
Fermi, E. 1949, Phys. Rev., {\bf 75} 1169

\noindent
Gaidos, J.A. et al. 1996, Nature, {\bf 383}, 319

\noindent 
Guerra, E.J. \& Daly, R.A. 1997, preprint, astro-ph/9707124

\noindent
Inoue, S. \& Takahara, F. 1996, ApJ, {\bf 463}, 555

\noindent
Kerrick, A.D. et al. 1995, ApJ, {\bf 438}, L59   

\noindent
Levinson, A. 1997, private communication.  

\noindent
Lin, Y.C., et al. 1992,  ApJ, {\bf 401}, L61

\noindent
Macomb, D.J., et al. 1995  ApJ, {\bf 449}, L99

\noindent
Mannheim, K. \& Bierman, P.L. 1992, A\&A, {\bf 253}, L21

\noindent
Maoz, D. 1997, in {\it Emission Lines In Active Galaxies}. eds.
B.M. Peterson et al., (SF Astr. Soc. of the Pacific), in press
\noindent

\noindent
Maraschi, L., et al. 1992, ApJ, {\bf 397}, L5

\noindent
Mattox, J.R., et al. 1997, ApJ, in press.

\noindent
Neuhoffer, G., et al. Phys. Lett. 1971, {\bf 37B}, 438   

\noindent
O'Dell, C.R., \& Wen, Z. 1994, ApJ, {\bf 436}, 194 

\noindent
O'Dell, C.R. \& Handron, K. D. 1996, Astr. J., {\bf 111}, 1630 

\noindent
Oort, J. H., 1950, Bull. Astr. Ins. Neth., {\bf 11}, 91

\noindent
Peterson, B.M., 1993, PASP, {\bf 105}, 247 

\noindent 
Punch, M. et al., 1992, Nature, {\bf 358}, 477

\noindent
Quinn, J., et al., 1996, {\bf 456}, L83

\noindent
Rybicki, G.B. \& Lightman, A.P. 1979, {\it Radiative Processes in Ap.}
(N.Y. Wiley)

\noindent
Schubnell, M.S., et al., 1996, ApJ, {\bf 460}, 644

\noindent
Schubnell, M.S., 1997, preprint, astro-ph/9707047

\noindent
Sikora, M., et al. 1994, ApJ, {\bf 421}, 153

\noindent
Stecker, F.W., et al. 1993, ApJ, {\bf 415}, L71

\noindent
Takahashi, T., et al. 1996, ApJ. (Lett) , in press   

\noindent
Thompson, D.J., et al. 1995,  ApJS, {\bf 101}, 259 

\noindent
von Montigny, C. et al.  1995  ApJ, {\bf 440}, 525
 
\vfill
\eject
 
\end{document}